\documentclass[twocolumn,showpacs,preprintnumbers,amsmath,amssymb]{revtex4}

\usepackage{graphicx}
\usepackage{dcolumn}
\usepackage{bm}

\begin{document}

\preprint{APS/123-QED}

\title{Scattering of surface and volume spin waves in a magnonic crystal}

\author{A. V. Chumak}

\email{chumak@physik.uni-kl.de}

\author{A. A. Serga}

\author{S. Wolff}

\author{B. Hillebrands}

\affiliation{Fachbereich Physik, Nano+Bio Center, and Forschungszentrum OPTIMAS, Technische
Universit\"at Kaiserslautern, 67663 Kaiserslautern, Germany}

\author{M. P. Kostylev}
\affiliation{School of Physics, University of Western Australia, Crawley, Western Australia 6009,
Australia}

\date{\today}

\begin{abstract}
The operational characteristics of a magnonic crystal, which was fabricated as an array of shallow
grooves etched on a surface of a magnetic film, were compared for magnetostatic surface spin waves
and backward volume magnetostatic spin waves. In both cases the formation of rejection frequency
bands was studied as a function of the grooves depth. It has been found that the rejection of the
volume wave is considerably larger than of the surface one. The influences of the nonreciprocity of
the surface spin waves as well as of the scattering of the lowest volume spin-wave mode into higher
thickness volume modes on the rejection efficiency are discussed.

\end{abstract}

\pacs{75.50.Gg, 75.30.Ds, 75.40.Gb}

\maketitle

Magnonic crystals, which are periodically structured magnetic materials, attract special attention
in view of their applicability for both fundamental research on linear and nonlinear wave dynamics
in artificial media, and for signal processing in the microwave frequency range \cite{MC review, MC
skyes, MC MSSW, Gubbiotti1, Gubbiotti2, Chumak_MC}. An array of parallel grooves formed on the
surface of a magnetic film seems to be one of the most effective methods to create a magnonic
crystal \cite{MC review, MC skyes, MC MSSW, Chumak_MC}.

Thin magnetic films support propagation of different types of spin-waves depending on the angle between
the wave propagation direction and the external magnetic field orientation. Backward volume
magnetostatic spin wave (BVMSW) and magnetostatic surface spin wave (MSSW) configurations are
characterized by parallel and perpendicular wave propagation relative to the magnetizing field applied
in the film plane \cite{Damon-Eshbach}. Both types of spin waves can be used in the magnonic crystal. As
a whole, MSSW devices offer more benefits for microwave applications in comparison to BVMSW devices, in
particular, because of more efficient excitation and reception by means of microwave antennas.
Furthermore, MSSW devices possess a noteworthy specificity: they are nonreciprocal, which means that
waves propagating in opposite directions in the film plane are localized at different film surfaces.
They also couple differently to surface scatterers (for example to microstrip antennas) which results in
nonreciprocal excitation of these waves \cite{PRB_Kostylev}. Thus, the investigation of propagation of
magnetostatic surface spin waves in an artificial magnonic crystal establishes the general problem of
propagation and scattering of nonreciprocal waves in structured media.

In our recent paper \cite{Chumak_MC} we presented results on scattering of BVMSW from a
quasi-one-dimensional periodic structure of grooves in a film surface. The main advantage of BVMSW-based
magnonic crystals was the excellent spin-wave signal rejection ratio of more than 30~dB. In contrast to
MSSW, BVMSW are reciprocal waves. In this work we compare the operational characteristics of MSSW-based
magnonic crystals with characteristics of BVMSW-based magnonic crystals.

\begin{figure}
\includegraphics[width=0.9\columnwidth]{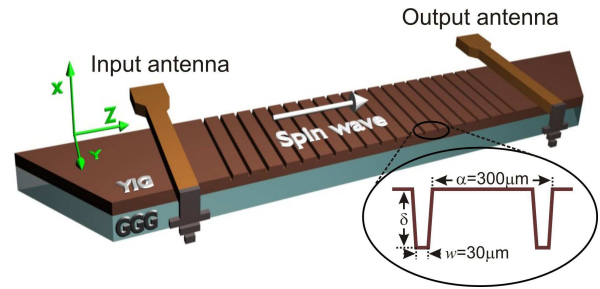}
\caption{\label{Setup} (Color online) Sketch of magnonic crystal structure used in the experiments.}
\end{figure}

To fabricate magnonic crystals, a $5.5~\mu$m thick yttrium iron garnet (YIG) film was used.
Photolithographic patterning followed by hot orthophosphoric acid etching was used to form the grooves.
The etch mask had 20 parallel lines of width $w=30~\mu$m spaced $270~\mu$m away from each other, so that
the lattice constant is $\alpha=300~\mu$m \cite{Chumak_MC} (see Fig.~\ref{Setup}). The grooves depth
$\delta$ was varied from 100 nm to $2.3~\mu$m by controlling the etching time and was measured using a
surface profilometer. Two microstrip antennas placed 8~mm apart on each side of the grooved area were
used to excite and receive spin waves as shown in Fig.~\ref{Setup}. A bias magnetic field of
$4\pi\cdot1.845$~A/m was applied in the plane of the YIG film, either along or perpendicular to the
$z$-axis depending on the type of spin waves under investigation. A microwave network analyzer was used
to measure transmission characteristics of these magnonic crystals.

\begin{figure}
\includegraphics[width=0.85\columnwidth]{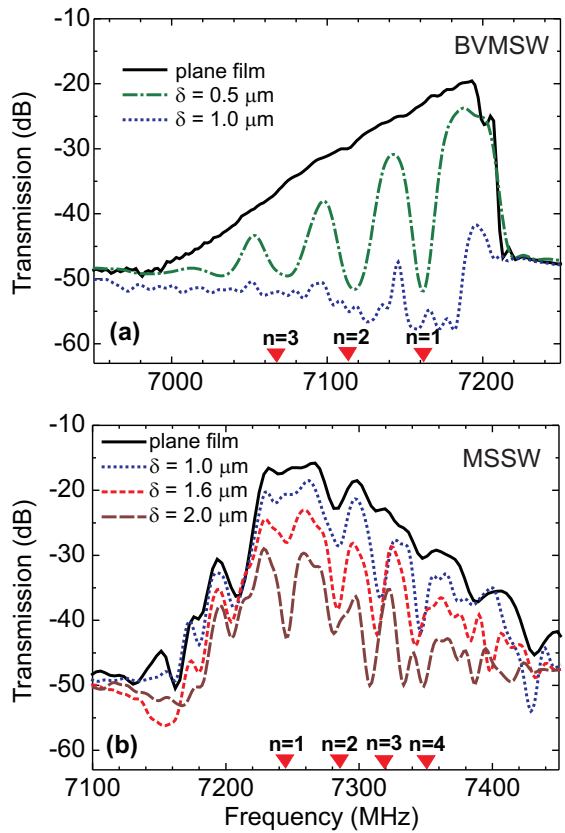}
\caption{\label{Transmission}
(Color online) Microwave transmission characteristics for an unstructured film (bold lines) and
for magnonic crystals measured for different groove depth $\delta$ in BVMSW (a) and MSSW (b) configurations.
Triangles show theoretically calculated positions of Bragg rejection bands of order $n$.
}
\end{figure}

The experimental BVMSW transmission characteristics for the unstructured film as well as for the
gratings with $\delta=500$ and 1000~nm are shown in Fig.~\ref{Transmission}(a). In this case the bias
magnetic field was applied along the $z$-axis (see Fig.~\ref{Setup}). The BVMSW transmission
characteristics for the unstructured film is limited from above by the ferromagnetic resonance frequency
and from below by a drop in the microwave antenna excitation efficiency for shorter wavelengths. The
insertion loss is determined by the energy transformation efficiency by the input and the output
antennas and by the spatial decay of spin waves. As can be seen in Fig.~\ref{Transmission}(a) grooves as
shallow as $\delta=500$~nm result in the appearance of a set of pronounced rejection bands (or
transmission gaps), where spin-wave transmission is highly reduced. The onset of the rejection bands
corresponds to a groove depth $\delta$ of 100~nm \cite{Chumak_MC}. For $\delta=1$~$\mu$m the insertion
loss in the whole spin-wave band is so pronounced that almost no spin-wave propagation is observed (see
Fig.~\ref{Transmission}(a)). Triangles in the figure show theoretically calculated positions of
rejection bands, with $n$ denoting the number of respective Bragg reflection band.

The measured MSSW transmission characteristics for the unstructured film as well as for the gratings
with $\delta$=1, 1.6, and 2~$\mu$m are shown in Fig.~\ref{Transmission}(b). In this case bias magnetic
field was applied perpendicular to the spin-wave propagation direction (i.e. along the $y$-axis in
Fig.~\ref{Setup}). The MSSW transmission characteristics for the unstructured film is limited by the
ferromagnetic resonance frequency from below and by the microwave antenna excitation efficiency from
above, thus it looks like a mirror reflection to the BVMSW transmission characteristics. Several
rejection bands in the transmission characteristics for the unstructured YIG film can be seen in
Fig.~\ref{Transmission}(b). These bands are formed due to ``exchange gaps'' in the MSSW spectrum. Their
origin is hybridization of MSSW with higher-order standing-wave resonances across the film thickness
\cite{Kalinikos86, PRL_restoration}. Positions and depths of these rejection bands are determined by the
film thickness and conditions for magnetization pinning at the film surface. Obviously, the potential
formation of exchange gaps is a drawback of the MSSW configuration.

Upon formation of the groove arrays on the surface of the YIG film new rejection bands appear (see
Fig.~\ref{Transmission}(b)). Some of them overlap the exchange gaps. A groove depth $\delta$=0.5~$\mu$m
(not shown in Fig.~\ref{Transmission}(b)) results only in a slight modification of the MSSW
transmission, while for BVMSW the same structure shows rejection bands of approximately 30~dB in depth
(see Fig.~\ref{Transmission}(a)). An increase in $\delta$ to 1~$\mu$m results in the appearance of
pronounced rejection bands (while for BVMSWs one sees complete rejection for this groove depth). Further
increase in $\delta$ results in an increase in the rejection efficiency and in insertion loss in the
pass bands. Thus, the operational characteristics of magnonic crystals with $\delta$=2.0~$\mu$m is
completely unsatisfactory.

Triangles in Fig.~\ref{Transmission}(b) show positions of the rejection bands calculated based on
Bragg's law. Apart from the formation of rejection bands two more effects occur. These effects are weak,
but noticeable. First, with increase in $\delta$ the rejection bands are slightly shifted towards lower
frequencies for the MSSW-based crystal. A similar shift but towards higher frequencies was previously
shown experimentally and theoretically for the BVMSW geometry \cite{Chumak_MC}. In both cases this shift
corresponds to a slight decrease in wave vectors of the waves, for which the Bragg condition is
fulfilled. Another effect, which seems to take place, is the increase in the depth of exchange gaps. We
suppose that it is caused by chemical processing of the film surface at the place of the grooves
\cite{etching}. Chemical etching increases magnetization pinning at the film surface which leads to an
increase in rejection band depth. Simultaneously the positions of these gaps are also slightly shifted
because the film thickness changes at the groove positions.

\begin{figure}
\includegraphics[width=0.9\columnwidth]{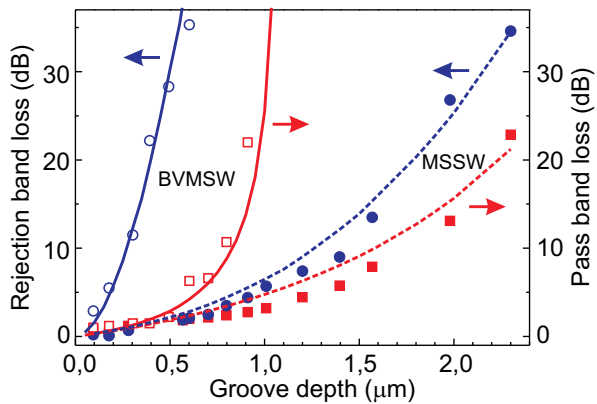}
\caption{\label{result} (Color online)
Insertion loss in the first-order rejection band (circles) and in the first pass band (squares)
measured for the BVMSW- and MSSW-based magnonic crystals (open and filled symbols, respectively)
as a function of the groove depth $\delta$. The solid and dashed lines present the insertion loss
calculated for BVMSW and MSSW cases, respectively.
}
\end{figure}

As seen in Fig.~\ref{Transmission}, the groove depth determines the transmission characteristics of the
fabricated magnonic crystals for both types of spin waves. In order to investigate this effect, the
insertion loss for the first-order rejection band and the parasitic loss in the first pass band for the
BVMSW and MSSW-based magnonic crystals are plotted in Fig.~\ref{result} as functions of the groove depth
$\delta$. The central frequency for the first-order gap for BVMSW is 7160\,MHz, and it is 7245\,MHz for
MSSW.

It can be seen from Fig.~\ref{result} that the behavior of the operational characteristics for MSSW and
BVMSW-based crystals is qualitatively similar: with increase in the groove depth the loss in the
rejection bands increases as well as the loss in the pass bands, but the latter effect is smaller.
However, one sees that for MSSW-based magnonic crystals the increase in loss is weaker, or, in other
words, MSSW is not as sensible to inhomogeneities at the film surface as BVMSW. For example, for a
5.5~$\mu$m-thick YIG film and for the BVMSW configuration the rejection efficiency reaches 30~dB for
$\delta \approx 0.5$~$\mu$m, while for the MSSW configuration this groove depth results in a rejection
about six hundred times smaller (2~dB).

The dependencies shown in Fig.~\ref{result} can be approximated by the simple model we previously
presented in Ref.~\cite{Chumak_MC}. This model is based on the analogy of a spin-wave waveguide
with a microwave transmission line. In the framework of this theory it was assumed that spin-wave
reflection is caused by periodical variation of the effective characteristic impedance of the
spin-wave waveguide due to the periodic variation of the YIG-film thickness in the grooved area. As
it is seen from Fig.~\ref{result} the theory agrees well with the experimental data for the
BVMSW-based magnonic crystal. However, in order to achieve the qualitative agreement with the
experimental data we had to multiply the theoretical reflection coefficient with an empirical
coefficient $\eta = \eta_\mathrm{BVMSW} = 6$ \cite{Chumak_MC}. For MSSW one also obtains good
qualitative agreement with the experiment, however to obtain quantitative agreement a much smaller
value for the empirical coefficient $\eta = \eta_\mathrm{MSSW} = 0.5$ is necessary. This
phenomenological theory does not reveal physical mechanisms underlying the drastic difference in
the spin-wave scattering efficiencies for these two types of waves. However, it suggests that the
behavior of both types of waves is qualitatively the same. In order to explain the difference in
rejection one has to explain the factor 12 in the strength of scattering from a single groove.

The problem of scattering of spin waves from inhomogeneities can be posed as a conventional
integral-equation formulation of the scattering problems \cite{scattering}. In the case of BVMSW
\cite{Chumak_MC} the integral equation is scalar, but for MSSW it is a vector one and makes use of the
whole tensorial Green's function of excitation of spin waves by an external source \cite{PRB_Kostylev}.
Both types of spin waves possess an in-plane and an out-of-plane component of the dipole field. However,
only the out-of-plane component of the dipole field exerts a torque on the magnetization vector in the
BVMSW case, while for MSSWs both field components contribute to the torque. The vector character of the
integral equation for MSSW reflects the latter fact. MSSW are non-reciprocal waves, which is also
reflected in the Green's function of excitation \cite{PRB_Kostylev}. However, solving the scattering
problem for the periodical potential of grooves in the first Born approximation \cite{scattering} gives
a result which is of the same form as was previously found for BVMSWs in Ref. \cite{MC_JAP}. No effect
of nonreciprocity is seen in the expressions derived for MSSW. Furthermore, in the first Born
approximation the amplitude of the wave transmitted through the groove structure is of the same order of
magnitude for both BVMSW and MSSW. This result is in clear contradiction with the experimental data.

Thus, accurate numerical modeling of both BVMSW and MSSW cases (which is out of the scope of this paper)
is necessary to explain the observed large difference in the depths of rejection bands for these two
cases. In contrast to Ref. \cite{scattering, PRB_Kostylev, MC_JAP}, this model should be
two-dimensional, i.e. it should include dynamic-field variations across the film thickness, since one of
the possible contributions to this difference is coupling of the incident BVMSW to higher-order BVMSW
thickness modes in the grooved area. The latter is seen in the thickness-resolved integral equation
formulation of the problem of BVMSW scattering. More efficient transfer of energy of the incident
lowest-order mode of BVMSW into the higher-order modes when the condition for the standing-wave
resonances (Bragg's law) in the grooved area is met, can be responsible for the increased depths of the
rejection bands in the BVMSW case.

In conclusion, we have experimentally demonstrated the strong difference of operational
characteristics of chemically etched one-dimensional magnonic crystals for cases when reciprocal
backward volume magnetostatic spin waves or nonreciprocal magnetostatic surface spin waves were
used as signal carriers. It has been shown that even small regular distortions of the surface of a
magnetic film result in the appearance of pronounced rejection bands in the BVMSW frequency
spectrum. At the same time such distortions only slightly affect the propagation of the surface
spin wave. The scattering of the lowest BVMSW mode to the higher-order thickness modes is assumed
as a possible mechanism of the observed effective rejection of this wave.

Financial support by the DFG SE 1771/1-1, Australian Research Council, and the University of
Western Australia is acknowledged.

\end{document}